\documentclass[12pt,preprint]{emulateapj}

\usepackage{natbib} 
\usepackage{epsfig} 
\usepackage{graphicx}  
\usepackage{rotating}

\begin{document}


\title{ON THE ANTHROPIC PRINCIPLE IN THE MULTIVERSE: ADDRESSING
PROVABILITY AND TAUTOLOGY}



\author{Douglas~F.~Watson} \affiliation{Department of Physics and
Astronomy, Vanderbilt University, Nashville, TN 37235}

\begin{abstract}\label{abstract}
In this \emph{Letter} we examine the Multiverse theory and how it
relates to the Anthropic Principle.  Under the supposition of Eternal
Inflation, the String Theory Landscape (STL) has reinvigorated the
discussion of the Anthropic Principle.  The main premise being that
the fundamental constants of our Universe are not necessarily of any
fundamental physical importance, rather that the specific values are
requisite for intelligent life to arise, and hence, for intelligent
life to measure such constants.  STL predicts a multitude of other
meta-stable Universes with fundamental constants different than our
own, possibly hinting at some intrinsic \emph{specialness} of human
life.  We develop a theoretical framework to prove whether, (1) the
Universe we observe must be consistent with the existence of
observers, (2) the principle is only ontological in nature, or (3) if
the Anthropic Principle itself is simply a tautology.\\


\end{abstract}



\section{INTRODUCTION}

Is there a peculiar specialness to our Universe with regard to the
existence of intelligent life?  Philosophers have
long pondered whether or not the Universe is ``fine tuned'' to
accomodate such conscious beings.  This is, in the most general sense,
what is known as \emph{The Anthropic Principle}.  It can be considered
in direct contradiction to the Copernican Principle (or the
Cosmological Principle) that postulates that man does not occupy a
privileged position in the Universe (under this principle, if we
observe that the Universe is isotropic then it can be easily shown
that the Universe is homogeneous).  More specifically, the Earth is
not the center of the Universe, and therefore,  by definition,
\emph{you} are not the center of the Universe.  This can also be
extended to the \emph{Mediocrity Principle} which
argues that there is nothing unique about the Earth or even us as
humans \citep[see][]{zaius68}.  Furthermore, it has also been
frequently proposed (e.g., Descartes) that the Anthropic
Principle is a circular (or tautological) argument, and, in addition,
since all other Universes in the Multiverse are inaccesible to us,
this discussion is therefore meaningless.  Until recently, most physicists 
would most likely shy away from addressing the Anthropic Principle in 
any scientific body of work.  However, recent cosmological unravellings 
have made this subject unavoidable. Thus, the fundamental question
to then pursue is: do we observe the Universe as it is simply because
only observers like ourselves exist in such a Universe?

Despite precision measurements from the Cosmic Microwave Background
and large scale structure (LSS) confirming the homogeneity of the
Universe, the Anthropic Principle is not necessarily invalidated.  A
consequence of both String Theory and Cosmic Inflation (more
specifically, Eternal Inflation) is the existence of an infinite
number of  possible Universes.  Given this ``Multiverse'' scenario, we
can redefine the problem.  It may be plausible to assume that the fundamental
constants need to have almost the exact values they do to give rise to
life as we know it.  Furthermore, cosmological paramaters, like
$\Omega$, need to be tuned to incredibly high precision to result in a
Universe that is not too ``diluted''  or too ``overweight''.  However,
the Multiverse implies that these constants will  be different for
every ``pocket universe''.  In this paper, we form, from first
principles, mathematical relations that attempt to infer the
probability that we as humans matter in the Multiverse.


\section{THEORETICAL FRAMEWORK}\label{model}  
In the following sections we lay out our theoretical framework for
calculating whether or not we are of any significance in a seemingly
void, indifferent, and desolate Universe(s).  We note that we do not
subdivide our investigation to consider more specific versions of the
Anthropic Principle, like the Weak\footnote[1]{We strongly advise not
using the word ``weak'' when coming up with a Principle.} and Strong
versions.

The Conditional Relations for the Anthropic Principle model (Model 1)
assumes the favored concordance $\Lambda$CDM paradigm, simplified
predictions from \emph{The String Theory Landscape} (STL), and
assumptions about the Multiverse from Eternal Inflation to calculate
the likelihood that we matter, which is not at all arrogant endeavor.
We now lay out a simple theoretical approach to estimate the magnitude
of the importantness of our intelligent species.

\begin{figure}[t]
\begin{center}
\includegraphics[width=.5\textwidth,height=.3\textheight]{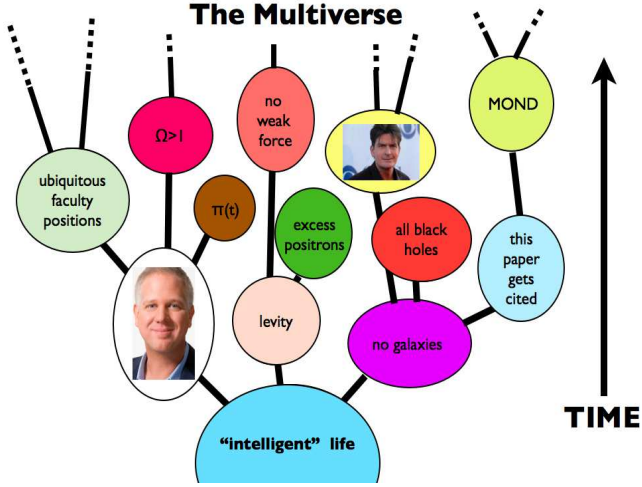}
\caption{Cartoon schematic of the Multiverse. We inhabit one of a
multitude of meta-stable Universes.  Eternal inflation predicts that
these ``pocket universes'' will have fundamental constants different
than ours. Hence, though there may be other forms of life, we do not
expect them to remotely resemble the intelligent life in our own
Universe. Image is to scale.}
\label{fig:fig1}
\end{center}
\end{figure}

We first take a standard envelope and turn it over.  We then start to
write down math on the back of it that we think relates to
probabilistic arguments relevant to the Anthropic Principle.  We will
assume that there are $\sim10^{200}$ meta-stable Universes comprising
the Multiverse allowed by STL\footnote[2]{This may or may not have
been ``borrowed'' from a recent talk by Alan Guth.}. Let us also
assume that the fundamental constants in each of these $\sim10^{200}$
Universes are different from ours, though their vacuum energy
densities may be somewhat similar.  Thus, life may certainly exist in
this multitude of Universes, but we argue that it will not be remotely
like ours (see Figure \ref{fig:fig1}).  We also now know, to great
accuracy, that our Universe is $\sim 6000$ years old ($\pm 13.7$
billion years), and let us take the upper limit.  Since humans have
been around $\sim 5999 \frac{360}{365}$ years, this requires that the
current epoch is really the only time that mankind could have
existed.

Mathematically, the probability that humans exist (denoted as
$m$), and therefore may matter, is :
\begin{equation}\label{eqn:1}
m = 10^{-200}\times 10^{-10} = 10^{-210}
\end{equation}
 This remarkably small value of $m$ lends creedence to the fact that
we may not be wholly irrelevant.

The Conditional Relations for the Anthropic Principle Proposed In the
Extended Regime model  (Model 2) pushes this argument even further.
This new regime is one in which we take knowledge of the specialness
of humankind and follow that semi-cogent path to uncover the
specialness of \emph{you}.  We know that $m$ is of order $10^{-210}$,
and now we fold in the probability that throughout all of humanity,
\emph{you} were created.  We start by adding the simplified conjecture
that every, and only, G-type stars have Earth-like planets that harbor
human life ($\sim 10^{10}$ in our own galaxy), and that there are some
$\sim 10^{10}$ MW-ish galaxies in our Universe.  Armed with the
knowledge that humans have been around $\sim 5999 \frac{360}{365}$
years (see the previous section for the detailed calculation), let us
postulate that over this time span there have been roughly $10^{10}$
human beings.  Your parents were 2 of these $10^{10}$, and there are
$\sim 4.2817\times 10^{22}$ genetically different zygotes for every
couple.  We define a new parameter $\epsilon$ with these
considerations in mind,

\begin{equation}\label{eqn:2}
\epsilon = \Big(\frac{1}{10^{20}}\Big) \times
\Big(\frac{2}{10^{10}}\Big) \times \Big(\frac{1}{4.2817\times
10^{22}}\Big),
\end{equation} 
What unfolds is quite remarkable, yet completely uncontrived. Close
examination of Eq. \ref{eqn:2} reveals that,
\begin{equation}\label{eqn:3}
\epsilon = 4.6709\times 10^{-53}
\end{equation}
\begin{equation}\label{eqn:4}
\epsilon = \frac{1.8439\times 10^{-55}}{4\pi ^{2}},
\end{equation}
which is just 42 times the Planck constant squared divided by
$4\pi ^{2}$:
\begin{equation}\label{eqn:5}
\epsilon = 42\times\frac{h^{2}}{4\pi ^{2}} = 42\hbar ^{2},
\end{equation}
Therefore, in units of $\hbar^{-2}$,
\begin{equation}\label{eqn:6}
\epsilon =  42 .
\end{equation}

It may also be worth considering the interesting possibility that
$\epsilon$ could evolve with time, making you even more special in the
past \citep[see][]{scherrer09}.  Of course, we need to combine
$\epsilon$ and $m$ to attempt to quantify your significance, thus we
introduce the parameter $u$.  There are surely other possibilities
that could be considered, and we just add these as higher order terms
$O$ (naturally the uncertainties in the assumed WMAP5 parameters are
encompassed by $O$). In other words,
\begin{equation}\label{eqn:7}
u = \epsilon \times m + O,
\end{equation}
(again, in units of $\hbar^{-2}$).  The higher order terms can
be set as an upper limit, thus we can re-write Eq. \ref{eqn:7}, which
now says that,
\begin{equation}\label{eqn:8}
u \leq m\epsilon, \ \textbf{ QED}.
\end{equation}
Our theoretical argument certaintly bolsters the Anthropic Principle.
Specifically, the models imply that we as self-aware beings are
special in the Multiverse, and possibly more importantly, that
\emph{you} are important, though $\leq m\epsilon$.

\section{The Acronym Cleverness - Number of Citations Relation}\label{discussion}
\begin{figure}[h]
\begin{center}
\includegraphics[width=.5\textwidth,height=.3\textheight]{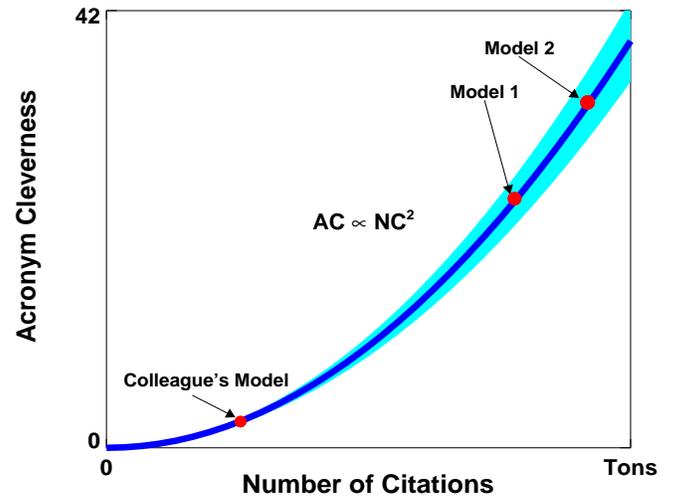}
\caption{\emph{AC-NC} diagram.  We project where this paper will lie
on the \emph{AC-NC} relation.  Solid band represents some sort of
subjectiveness.  Regardless of robustness, the weaker acronym for
Model 1 has the potential to make our paper less relevant.  We also
show an anonymous colleague's model for comparison.  See \S
\ref{model} for acronyms and details.  }
\label{fig:fig2}
\end{center}
\end{figure}

While our rigorous mathematical proof in \S \ref{model} appears
infallible, Fig. \ref{fig:fig2} mandates that we proceed with caution,
and that only time will tell how our theory holds up.  In this figure
we project where this paper will fall on the AC-NC diagram.  

By standard convention, we investigate the power of our acronyms by
making use of the Great Ramification Evaluator of Acronyms Tester
\citep{jennings_rutter11} -- a modified version of the W\"URST test
\citep{germans56}.  Since $AC \propto NC^{2}$, Model 1 has the
potential to drive down the number of citations, and cause the paper
(and hence our conjectures) to lose credibility.  In fact, the
previously published paper of the Probabilistic Evoking Of
Non-essentiallity model (a colleague's model, not ours) actually uses
our exact proof, essentially in reverse, to show that all of this
conspiring actually proves that we are incredibly insignificant in
what they dub ``the grand scheme of things''.  Despite being far
superior to our models and strongly suggesting our total lack of
importance, their poor choice of acronym causes them to have an
incredibly low citation count rendering their theory irrelevant.
Some\footnote[3]{Actually, no one} have argued that the $AC-NC$
relation may be violated.  To address this concern, in a forthcoming
paper \citep[][In Prep.]{watson11} we develop the Conditional
Relations for the Anthropic Principle for Probabilistic Informatic
Extreme Set Theory model, which actually is only an acronym (a
particularly clever one, and replete with big words) and should
potentially receive no citations, hence, violating the laws of Figure
\ref{fig:fig2}.  Some concrete examples of the power of the AC-NC
relation include BOSS \citep{BOSS} and SHAM \citep{SHAM}.  However,
possible serious violations to consider are 2D-FRUTTI \citep{2DFRUTTI}
and PROSAC \citep{PROSAC} \citep[see
also;][]{COYOTES,WOMBAT,SAURON,COLA,FLAMINGOS,BICEP,TOOT,CREAM,GADZOOKS,MARTINI,LasDamas}.

We conclude that time will tell whether or not you matter, however
further consideration of the blatant circularity of this entire
argument needs to be investigated, which has the potential to render
this entire paper ipso facto meaningless.

\section{ACKNOWLEDGEMENTS}

Assuming no permanent blacklisting, DFW would like to thank any future
employer.  DFW is funded by his adviser in exchange for a modest
publication rate along with superlative and punctual morning coffee.

\bibliography{aprilfools.bib}

\end{document}